\documentclass[letterpaper]{appolb}
\usepackage{graphicx}

\begin{document}

\title{Strangeness and the Quark-Gluon Plasma:\\
Thirty Years of Discovery
\thanks{Presented at SQM2011, Krak\'ow, Poland}
}
\author{Berndt M\"uller
\address{Department of Physics, Duke University, Durham, NC 27708, USA}
}
\date{\today}
\maketitle

\begin{abstract}
I review some aspects of the role of strange quarks in hot QCD matter and as probes of quark deconfinement at high temperature.
\end{abstract}
  
\section{The Quark-Gluon Plasma: A Strange State of Matter}

Strange quarks play a crucial role in shaping the phase diagram of QCD matter (see Fig.~\ref{fig:BM1}):
\begin{figure}[htb]
\centering
\includegraphics[width=0.6\linewidth]{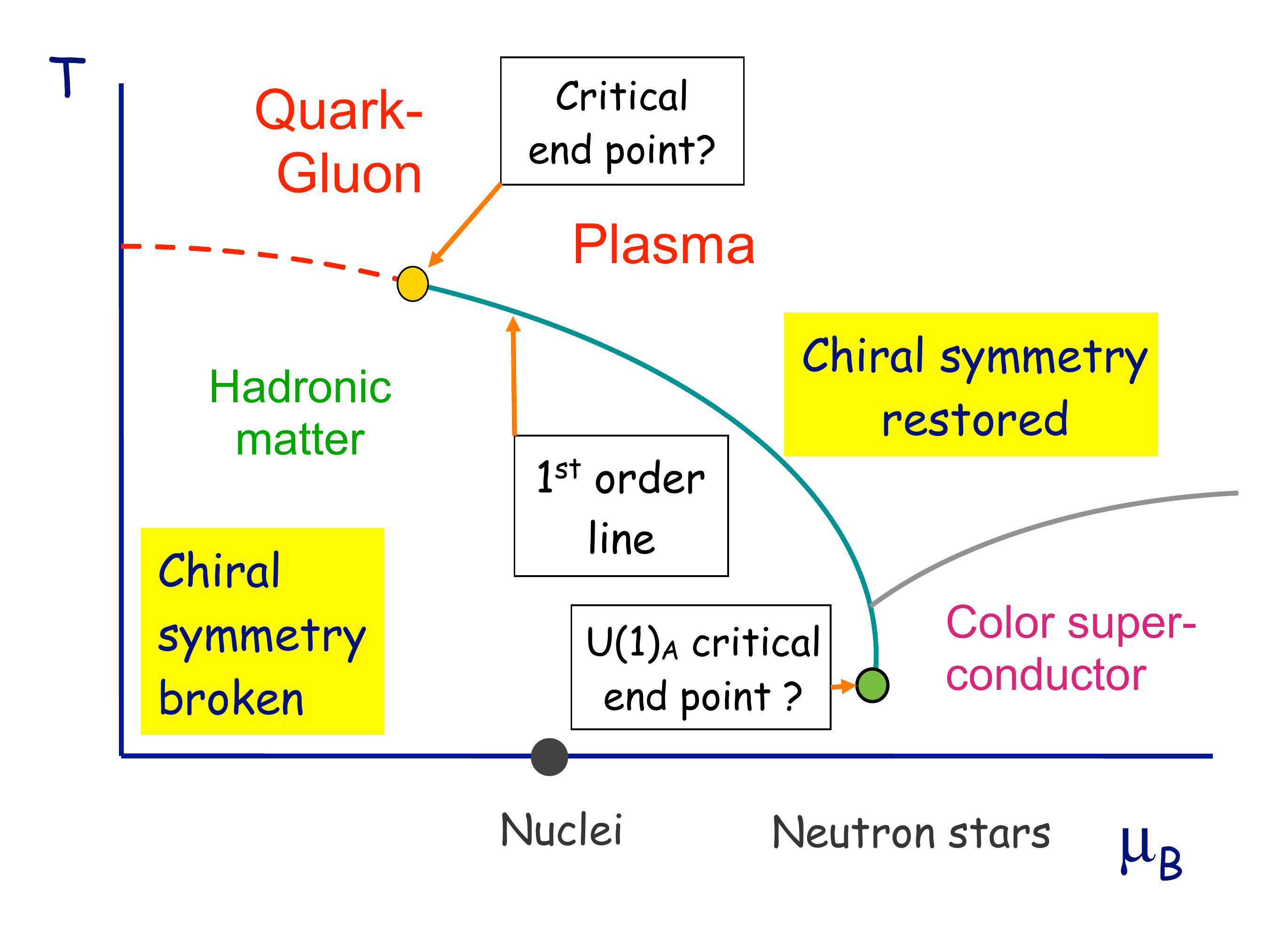}
\caption{Structure of the QCD phase diagram. It is important to keep in mind that the phase diagram assumed perfect thermal equilibrium conditions. The matter produced in relativistic heavy ion conditions can never be in complete equilibrium because of the rapid expansion.}
\label{fig:BM1}
\end{figure}
\begin{itemize}
\item The mass $m_s$ of the strange quark controls the nature of the chiral and deconfinement transition in baryon symmetric QCD matter \cite{Brown:1990ev}. As a consequence, $m_s$ also influences the position of the critical point in the QCD phase diagram, if one exists.
\item The mass of the strange quark has an important effect on the stability limit of neutron stars and on the possible existence of a quark core in collapsed stars \cite{Alford:2004pf}.
\item Strange quarks enable the formation of a color-flavor locked, color superconducting quark matter phase at high baryon chemical potential and low temperature by facilitating the symmetry breaking ${\rm SU}(3)_{\rm F} \times {\rm SU}(3)_{\rm C} \to {\rm SU}(3)_{\rm F+C}$ \cite{Alford:1998mk}.
\end{itemize}

Strange quarks are also excellent probes of the structure of QCD matter because:
\begin{itemize}
\item they are hard to produce at temperatures below $T_c$ since their effective mass is much larger than $T_c$ when chiral symmetry is broken, but easy to produce at temperatures above $T_c$ since the current mass of the strange quark $m_s \approx 100~{\rm MeV} < T_c$;
\item quark flavor is conserved under the strong interactions implying that strange quarks, once produced, are not easy to destroy during the dilute hadronic freeze-out stage of a heavy ion reaction.
\end{itemize}
\begin{figure}[htb]
\centering
\includegraphics[width=0.45\linewidth]{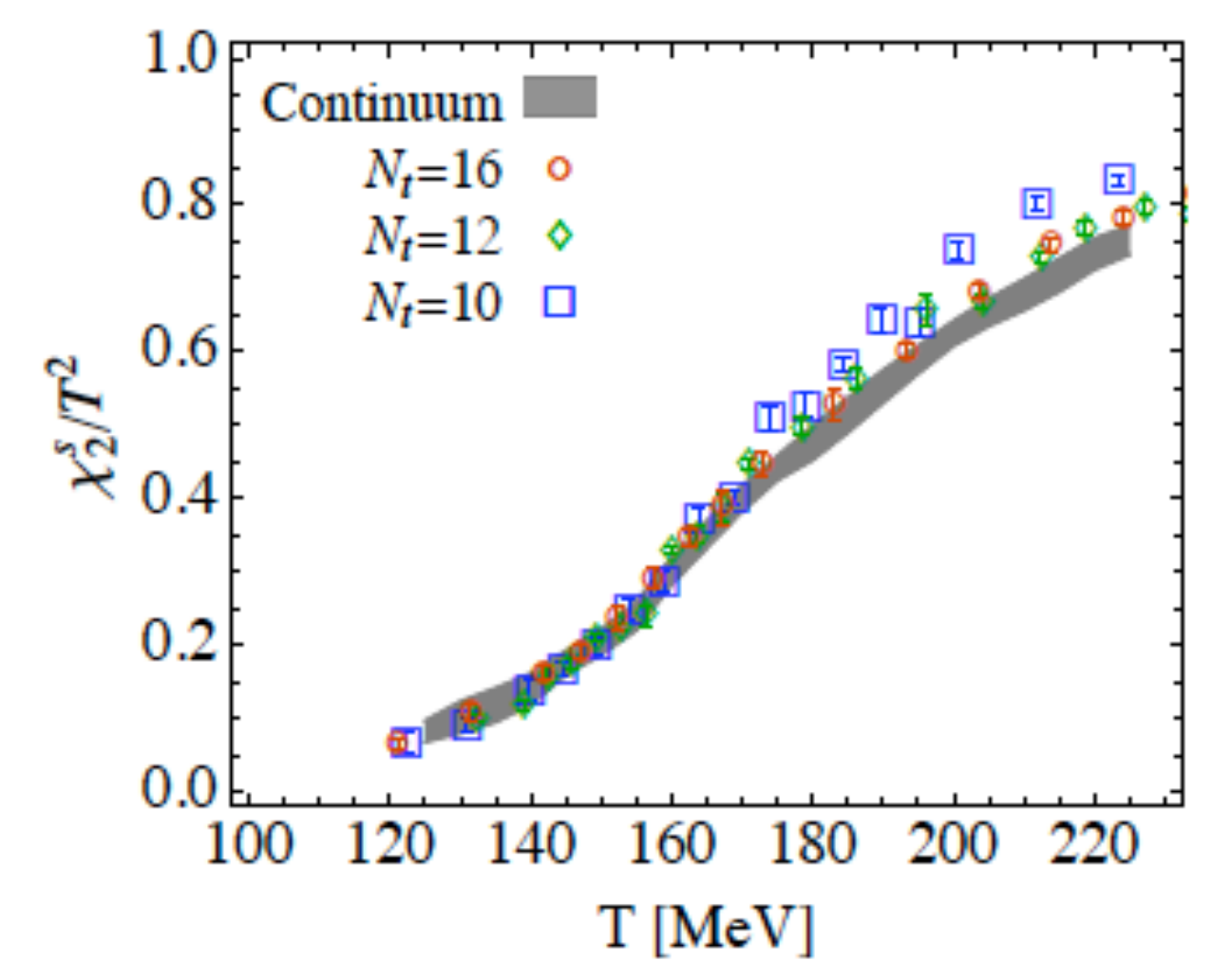}
\includegraphics[width=0.53\linewidth]{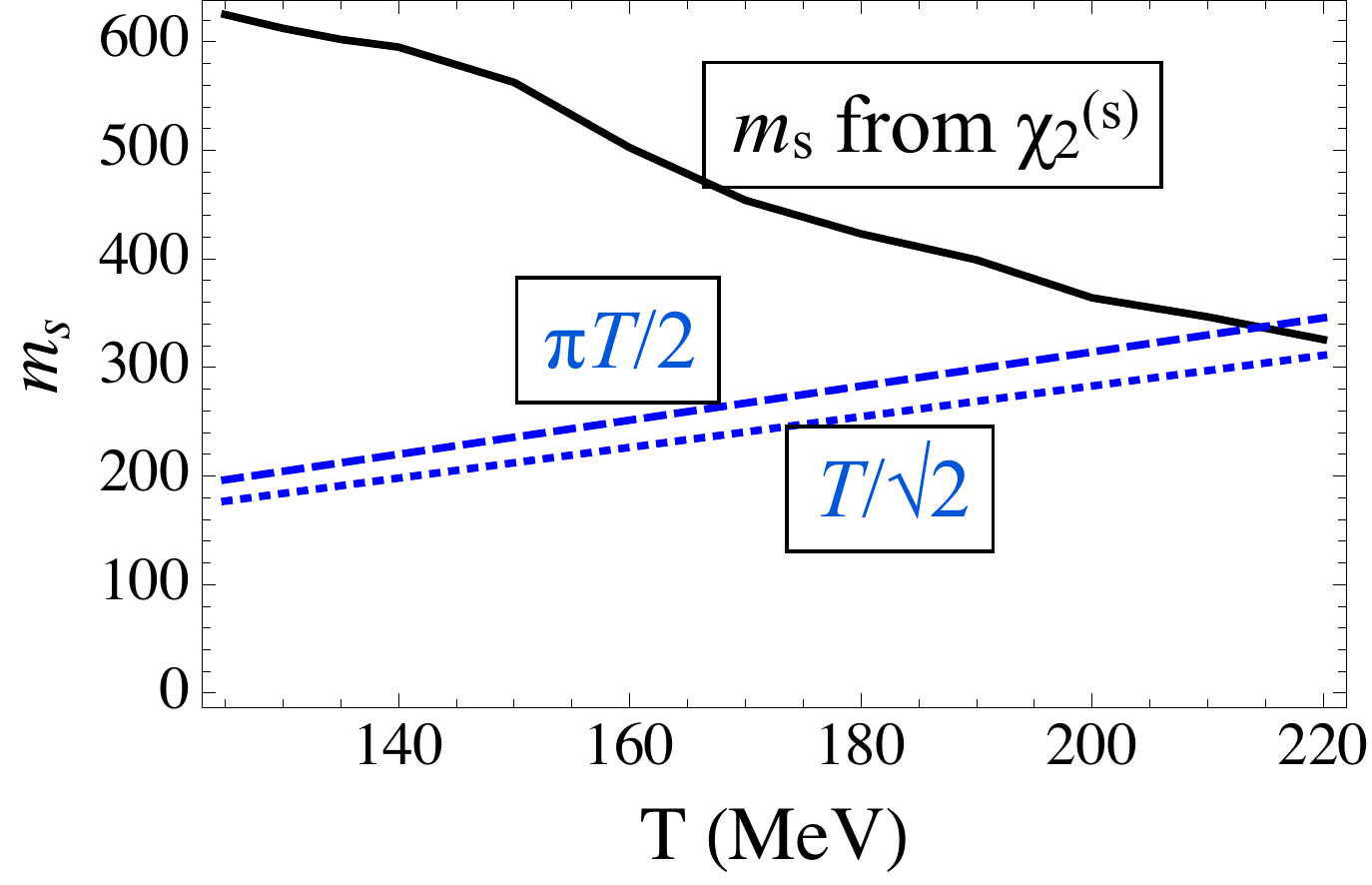}
\caption{Left panel: The inflection point of the strange quark number susceptibility $\chi_2^s/T^2$ serves as an indicator of the location of the QCD deconfinement transition. Right panel: Effective mass of the strange quark as a function of temperature, in comparison with typical thermal fermion mass scales at weak and strong coupling.}
\label{fig:BM2}
\end{figure}
In view of these considerations, the strange quark susceptibility $\chi_s$ is considered as a signature of the deconfinement transition in lattice QCD \cite{Aoki:2006br,Borsanyi:2010bp}. As Fig.~\ref{fig:BM2} (left panel) shows, the ratio $\chi_s/T^2$ grows rapidly, but smoothly across $T_c$. If one interprets the rise as a change in the effective mass of the strange quark, the change reflects a drop from typical values of the constituent strange quark mass ($M_s \sim 500-600$ MeV) to thermal quark masses (right panel). Off-diagonal susceptibilities of conserved quantum numbers including strangeness, such as $C_{BS}$ and $C_{QS}$, also can serve as indicators of the transition from hadronic matter to deconfined matter at $T_c$ \cite{Koch:2005vg,Majumder:2006nq}.

\section{Strangeness Enhancement}

The original idea of enhanced production of hadrons containing strange quarks as a signature of quark deconfinement, proposed by Rafelski and Hagedorn in 1980 \cite{Rafelski:1980rk}, was based on the insight that strange antiquarks are at least as abundant as light antiquarks in quark matter for $\mu_B>0$, and can be substantially more abundant in ``many cases of interest'', e.g.\ when $\mu_B \sim T$ in the domain of collision energies accessible at the CERN-SPS. It is also worth noting that this original publication does not mention the multi-strange antibaryons $\overline{\Xi}, \overline {\Omega}$ and the $\phi$-meson, probably because these were considered too exotic and rare to be studied. But less than two years later, Rafelski discussed these probes prominently \cite{Rafelski:1982ii}, explaining that the formation of a baryon-rich, flavor equilibrated quark gluon plasma would result in their copious production.\footnote{This prediction was greeted almost with disbelief by many physicists in view of the fact that these particles are extremely rarely produced in $p+p$ collisions at SPS.}

The missing piece of the argument, that the flavor composition of a transient quark-gluon plasma could be equilibrated during the short period of its existence, was supplied in 1982 by our calculation \cite{Rafelski:1982pu} of the rate of strange quark pair production in a deconfined QCD plasma. The crucial observation was that the process $gg \to s\bar{s}$ is the dominant production channel. Thus an enhanced production of hadrons containing multiple strange quarks in nucleus-nucleus collisions, compared with $p+p$ collisions, not only signals quark deconfinement, but also the liberation of gluons as dynamical excitation modes in hot QCD matter. In other words, {\em strangeness enhancement}, as the effect was called in brief, provides a signal for the formation of a quark-gluon plasma in the true sense of the term.

The history of this insight has been described in great detail by Rafelski in the Proceedings of the Zimanyi Memorial Conference \cite{Rafelski:2007ti}; the interested reader is referred to this authoritative account. The full set of experimental implications of enhanced strange quark production were worked out in 1985 and published in a {\em Physics Reports} article by Koch {\em et al.} \cite{Koch:1986ud}, which made two main predictions. As a consequence of quark-gluon plasma formation in relativistic heavy-ion collisions:
\begin{enumerate}
\item The strange quark phase space will be fully equilibrated (in commonly used notation: $\gamma_s \ge 1$) at collision energies where most of the volume is converted into a quark-gluon plasma existing significantly longer than 1 fm/c. In fact, if strange quarks reach equilibrium abundance at a temperature significantly above $T_c$, they may become over-abundant ($\gamma_s > 1$) as the matter cools, and hadronization may occur under conditions of strangeness supersaturation \cite{Letessier:2006wn,Petran:2011aa}.
\item {\bf All} strange hadrons, including multi-strange baryons and antibaryons, will be produced in {\em relative} chemical equilibrium abundances. This is a natural consequence of hadron formation by statistical coalescence of deconfined quarks. 
\end{enumerate}
Especially the second prediction seemed preposterous at the time. However, it is fair to state that of all the quark-gluon plasma signatures proposed in the 1980's, equilibrated strangeness production is the only one which has been quantitatively confirmed in every detail by the experimental data. The predicted enhancement of strange (anti-)baryon production has been observed in collisions of heavy nuclei (Pb+Pb or Au+Au) at SPS, RHIC, and most recently, LHC, as chronicled in B. Hippolyte's lecture at this conference.

\section{Hadronization Mechanisms}

In order to make quantitative predictions for the expected hadron abundances in the final state of a relativistic heavy ion reaction it is important to understand how the quark-gluon plasma hadronizes. Rafelski and Danos addressed this issue in 1987 \cite{Rafelski:1987un}. They distinguished between fragmentation and recombination mechanisms and pointed out that recombination is expected to be the dominant mechanism of hadron emission at energies above the mean thermal energy. In this domain, the enhanced strangeness  production would most directly reflect the strange quark abundance and quark deconfinement in the quark-gluon plasma phase. 

\begin{figure}[htb]
\centering
\includegraphics[width=0.46\linewidth]{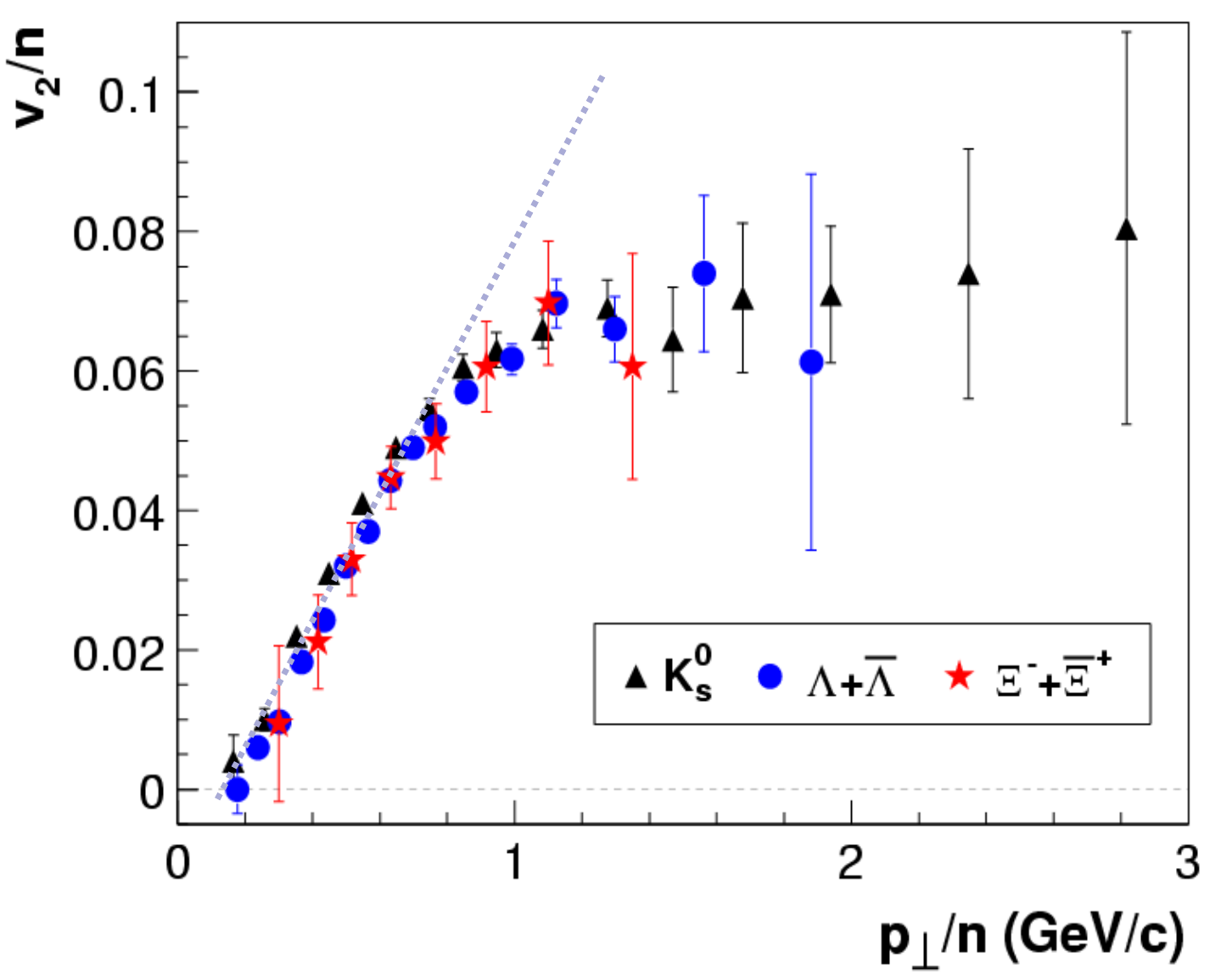}
\hspace{0.02\linewidth}
\includegraphics[width=0.5\linewidth]{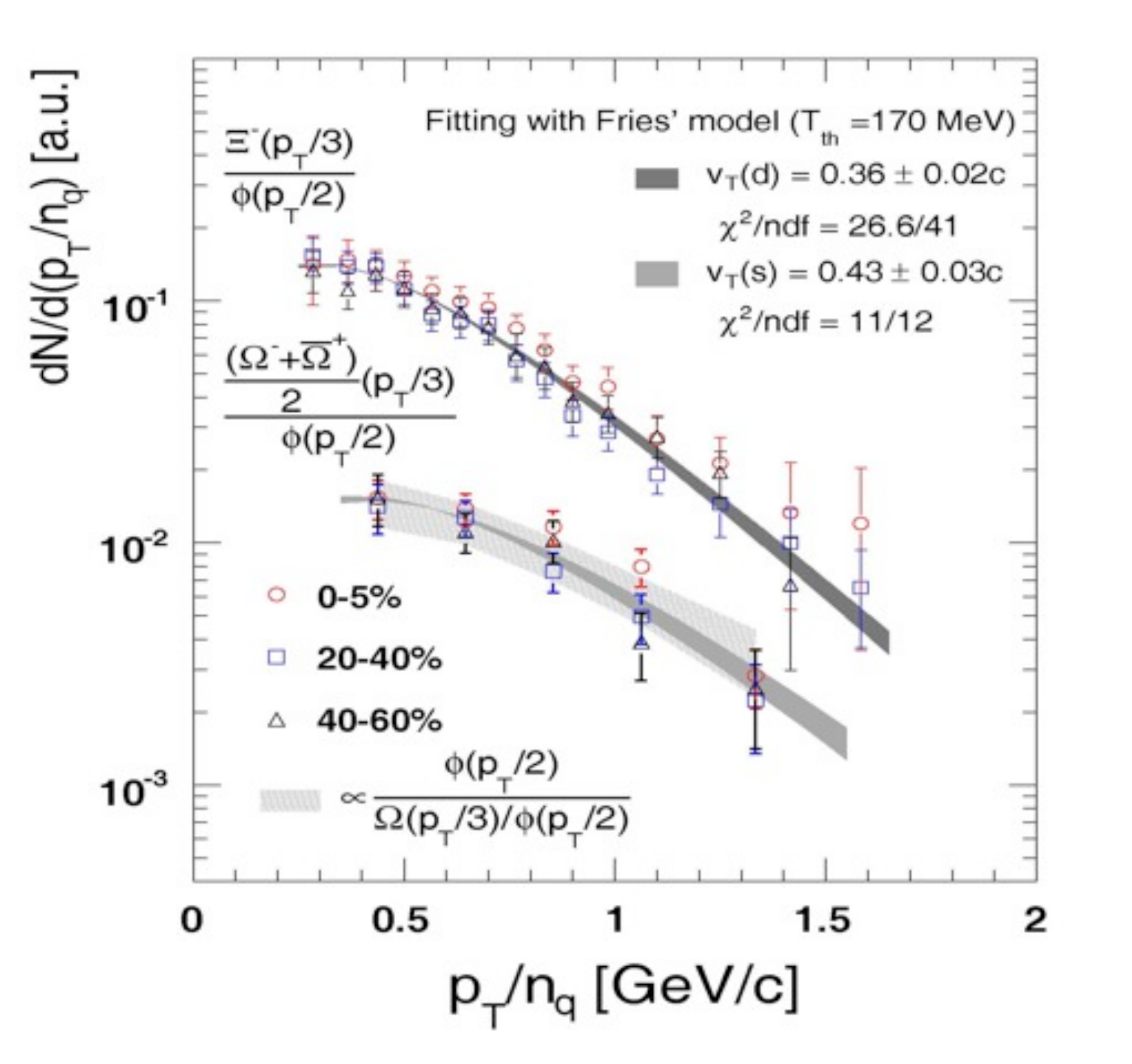}
\caption{Left panel: Elliptic flow velocity $v_2(p_T)/n_q$ per valence quark for several hadrons containing strange quarks. The data are from the STAR collaboration. Right panel: Quark spectra at hadronization deduced from ratios of hadron spectra in the framework of the valence quark recombination model \cite{Chen:2009zzo}.}
\label{fig:BM3}
\end{figure}

This picture has been confirmed by the measurements of relative hadron abundances and angular distributions in heavy ion collisions at RHIC, where quark recombination is believed to the source of the enhance baryon-to-meson ratio at transverse momenta in the $1-3$ GeV range and to explain the scaling of the elliptic flow velocity with valence quark number \cite{Fries:2003kq,Fries:2003vb,Greco:2003xt,Greco:2003mm} (see Fig.~\ref{fig:BM3}, left panel). The insight that the spectra of multi-strange hadrons (including the $\phi$-meson) rather faithfully reflect the primordial spectrum of the quarks from which they are formed by recombination can be used to infer the quark spectrum in the quark-gluon plasma just before hadronization. For example:
\begin{eqnarray}
N_s(p_T) &\sim& \frac{N_\Omega(p_T/3)}{N_\phi(p_T/2)} \sim \frac{[N_\phi(p_T/2)]^2}{N_\Omega(p_T/3)} ;
\\
N_d(p_T) &\sim& \frac{N_{\Xi^-}(p_T/3)}{N_\phi(p_T/2)} .
\end{eqnarray}
Quark spectra deduced in this manner by Chen {\em et al.} \cite{Chen:2009zzo} are shown in Fig.~\ref{fig:BM3} (right panel).

Whether hadronization is a sudden process, not only for hadrons emitted at high speed, but for hadrons in the bulk, is still unclear. Supercooling of the plasma phase could provide for such a mechanism, but lattice QCD calculations do not suggest that the deconfinement transition is of first order, at least not at $\mu_B \ll T$.  Another possible mechanism for sudden hadronization is a hydrodynamical instability near $T_c$ induced by a peak in the bulk viscosity, which could result in sudden cavitation of the quark-gluon plasma at the transition to hadronic matter \cite{Rajagopal:2009yw}.

The hadronization mechanism has implications for the expected integrated hadron yields. As pointed out by Rafelski and Letessier \cite{Rafelski:2000by}, if the hadronization of a quark-gluon plasma is an explosive process, the hadron yields should reflect rather faithfully the quark abundances before hadronization. On the other hand, if hadronization proceeds gradually allowing for the chemical equilibration of hadron abundances during hadronization, the hadron yields should reflect nearly perfect hadrochemical equilibrium \cite{Andronic:2009qf}. 

There seems to be general agreement that the yields of different hadrons are well described by the statistical model with a small set of adjustable parameters. The most important ones are the chemical temperature $T_{\rm ch}$, the baryon chemical potential $\mu_B$, and the strange phase space saturation factor $\gamma_s$. The chemical potential of the strange quark, $\mu_s$, is fixed by the requirement that the net strangeness of the matter must vanish. In the canonical formalism, applied to a finite matter volume, the fugacity factor $\gamma_s$ receives a contribution from strangeness conservation. Thus $\gamma_s$ is predicted to be less than unity in a finite volume even at full chemical equilibrium \cite{Hamieh:2000tk}.  However, as mentioned above, $\gamma_s$ can exceed unity at hadronization, if strange quarks are chemically equilibrated at $T \gg T_c$ and the annihilation reactions cannot keep up with the cooling rate of the expanding quark-gluon plasma \cite{Letessier:2006wn}. 

Rafelski and Letessier \cite{Rafelski:2002jd} have argued that an explosively disassembling quark-gluon plasma might even result in an overpopulation of light quarks after hadronization, expressed as a fugacity $\gamma_q>1$, to account for the entropy carried by gluons in the plasma phase \cite{Kuznetsova:2006bh}. An alternative way of denoting such an overpopulation is to introduce a pion chemical potential $\mu_\pi$. The threshold of pion condensation, $\mu_\pi = m_\pi$ corresponds to $\gamma_q \approx 1.6$. It is worth noting that a nonzero value of $\mu_\pi$ can also arise in the late expansion stage of a hadronic gas because of the kinetic suppression of pion number changing reactions at temperatures $T \ll T_c$ \cite{Bebie:1991ij}.
 
The question which of these scenarios discussed above is realized in nature remains open. One issue is, for example, whether the yield of unstable hadrons such as the $K^*$ is included in the statistical analysis at face value, because it might be altered by filial state interactions. Hopefully, the heavy ion data from the LHC and the low-energy RHIC runs will be able resolve this debate. The surprisingly low $p/\pi$ ratio seen in the LHC data may be an indication that the strict chemical equilibrium model with $\gamma_q=1$ does not work. 

\section{Hadron Resonances}

\begin{figure}[htb]
\centering
\includegraphics[width=0.6\linewidth]{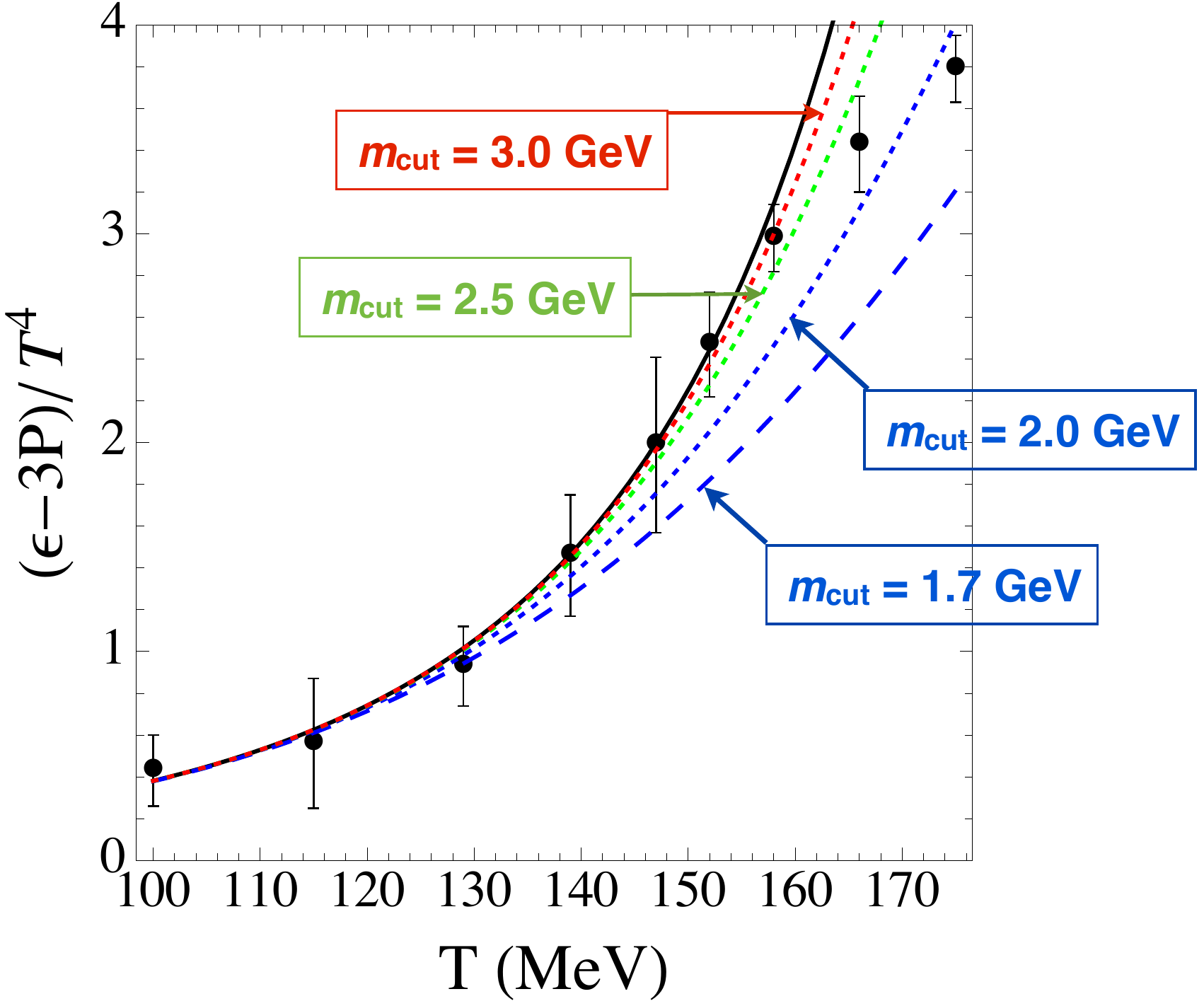}
\caption{QCD interaction measure asa  function of temperature $T$. The black dots (with error bars) show the results of a recent lattice calculation \cite{Borsanyi:2010cj}. The lines show the predictions for an exponential hadron mass spectrum with different mass cut-offs \cite{Majumder:2010ik}. The solid (black) curve includes all masses; the colored curves are for mass cut-offs $m_{\rm cut}$ as indicated in the figure.}
\label{fig:BM4}
\end{figure}

The calculation of hadron yields at chemical freeze-out temperatures near $T_c$ relies on the knowledge of the spectrum of excited hadrons. Usually, one takes the established hadron states listed by the Particle Data Group (PDG), but this may underestimate the hadron spectrum. Recently, lattice QCD simulations with physical quark masses have become sufficiently precise to shed some light n this question. The interaction measure $I = (\varepsilon-3P)/T^4$ is especially sensitive to the spectral density of excited hadrons, because the contribution of a state with mass $m$ is proportional to $m^2$. A recent lattice QCD calculation clearly shows an excess of the interaction measure above the predictions of the thermal model with the PDG states in the temperature range $140-170$ MeV  \cite{Borsanyi:2010cj}.  This excess can be explained \cite{Majumder:2010ik} if one assumes that the physical hadron spectrum grows exponentially as anticipated by Hagedorn \cite{Hagedorn:1967xx} (see Fig.~\ref{fig:BM4}). 

The presence of these ``Hagedorn states'' may contribute significantly to the chemical equilibration among hadrons in the temperature range near $T_c$ \cite{Greiner:2004vm,Greiner:2011rx}. It remains a challenge for the future to unambiguously distinguish between a scenario, in which the chemical composition of emitted hadrons reflects the suddenly frozen composition of the quark-gluon plasma, and one in which the chemical freeze-out temperature reflects the rapid disappearance of these heavy hadron resonances below a certain temperature. In reality, this apparent dichotomy may be an artificial consequence of our ignorance of the nature of the structure of strongly interacting matter near $T_c$. Since the transition between hadronic matter and quark-gluon plasma is not a discontinuous phase transition but a cross-over, it is possible that QCD matter in the transition regime may be equally well described as a dense gas of Hagedorn resonances or as a strongly coupled plasma of quarks and gluons. 

This picture suggests an intriguing possibility. The exponential hadron mass spectrum includes a large fraction of unknown hadron states, many of which may be hybrid states contain excited glue. Such a scenario would explain where the entropy carried by gluons in the quark-gluon plasma goes when the plasma hadronizes: It simply becomes part of the entropy carried by hybrid hadrons! These hybrid hadron resonances will decay rapidly and in the process generate a large numbers of mesons containing light quark pairs, which contribute to the light quark fugacity $\gamma_q$.

\section{Outlook}

Experiments at the SPS, RHIC, and the LHC have shown that the phase space of strangeness is fully equilibrated in ultrarelativistic  heavy ion collisions, as was theoretically predicted 30 years ago. Formation of deconfined QCD matter, the quark-gluon plasma, is the most natural mechanism for strangeness equilibration. A quantitative understanding of the experimental data raises many questions about the quark-gluon plasma-to-hadron gas transition: Is it a sudden hadronization or a near equilibrium transition? What happens to the excited glue as the matter cools below $T_c$? What is the role of the unobserved, high-lying hadron resonances? How do late-stage hadronic reactions influence the conclusions of analyses that assume chemical freeze-out at a fixed temperature? The superb data coming from RHIC and LHC invite improved theoretical studies. The future of this field clearly lies in theoretical collaborations, which can bring all theoretical tools to bear on the difficult, but intriguing problems whose solution promise deep theoretical insights into the workings of QCD.
\\

{\em Acknowledgment:} This work was supported in part by a grant (DE-FG02-05ER41367) from the Office of Science. I thank Jan Rafelski for many fruitful collaborations, whose results are reviewed in this lecture, and for sharing his insights into the physics of hot QCD matter on many occasions.

\end{document}